\documentstyle[art12]{article}
\begin{document}

\begin{center}
{\bf HEAVY QUARK FRAGMENTATION INTO BARYONS    \\
IN A QUARK-DIQUARK MODEL} \\[4mm]
A.P.\,Martynenko, V.A.\,Saleev\\[4mm]
Samara State University, Samara, Russia

\end{center}

\begin{abstract}
In the framework of the nonrelativistic QCD and a quark-diquark
model of baryons we have obtained the fragmentation functions for heavy
quark to split into spin-1/2 and spin-3/2 double heavy	baryons.
It was predicted the production rates as well as the shape of the
energy spectra for the $cc-$ and $bc-$baryons in the region of
$Z^o$ peak at LEP collider.
\end{abstract}

\section{Introduction}

In the last decade the great success was obtained in the study of the
heavy quarkonium production and decay (see, for example,
Refs.\cite{1,1b}). The investigation of the processes with the heavy
quarks is based on the factorization hypothesis \cite{2}. The mass of
heavy quark $m_Q$ is much larger than the scale of strong interactions
$\Lambda_{QCD}$. So the presence of the small parameter
$\Lambda_{QCD}/m_Q$ has allowed one to separate the effects of small
and long distances. The heavy hadron production amplitude may be
presented as the product of the partonic part, which can be calculated
using the perturbative QCD, and the nonperturbative factor, which
describes the free quarks to final hadron transition. In the framework
of the nonrelativistic quark model, this nonperturbative part can be
presented through the quarkonium wave function at the origin
$\Psi(0)$, which is calculated using the potential method (see
\cite{3} and References therein). So, the approach based on PQCD and
nonrelativistic quark model is known as a nonrelativistic QCD (NR QCD)
\cite{4}.

The heavy hadron production via fragmentation prevails in the large
transverse momentum region at $e^+e^-$ and hadron colliders \cite{5}.
Recently it was shown \cite{4} that the fragmentation functions
$D_{Q\to M}(z,\mu)$ and $D_{g\to M}(z,\mu)$ for quark and gluon to
split into a heavy quarkonium with fraction $z$ can be calculated
within the framework of NR QCD. The fragmentation functions are the
process independent and can be applied to the $e^+e^-$, photonic and
hadronic production of heavy quarkonia. Here we demonstrate that the
NR QCD approach can be used for the calculation of the fragmentation
function $D_{Q\to B}(z,\mu)$ for heavy quark to split into double
heavy baryon .

The estimation of the production rates for baryons containing two
heavy quarks was done recently on the basis of the quark-hadron
duality \cite{6} as well as on the basis of the PQCD \cite{7}. In
spite of the some differences, it was suggested in Refs. \cite{6,7},
that the double heavy baryon production has two step. In the first
stage, there is the heavy quark $Q$ fragmentation into double heavy
diquark $(QQ)$ in the colour octet state. The second step consists in
the nonperturbative diquark fragmentation into a ($QQq$) baryon. It
was mentioned in Ref. \cite{7}, that this mechanism of the fragmentation
can be factorized into short and long distance contributions as the
same in Ref. \cite{4}. The normalization of the fragmentation function
$D_{Q\to B}(z,\mu)$ is determined by the model dependent value of the
$(QQ)$ diquark wave function at the origin $\Psi_{(QQ)}(0)$ as well as
by the additional suggestion about nonperturbative diquark
fragmentation mechanism into baryon. Note, that the calculation of the
$\Psi(0)$ for the colour object using potential model isn't grounded
and destroys the factorization of the long and short distance effects.
It was also suggested in Ref. \cite{7}, that the probability of the
diquark $(QQ)$ fragmentation into baryon is equal to unity and the
heavy diquark carries all of the momentum of the baryon $D_{(QQ)\to
B}(z,\mu)\sim \delta (1-z)$. These circumstances raise too high the
predicted values of the double heavy baryon production rates.

The model of the quark-hadron duality, which was used in Ref.\cite{6} for
the prediction of $cc-$baryon production cross section at $B-$factory,
gives us the opportunity to obtain only upper limit of the production
rates and doesn't
predict baryon spectra. It was suggested in \cite{7} that all
($cc$)-pairs in the colour octet state with invariant mass from $2m_c$
to $2M_D+\Delta M$ ($M_D$ is $D$-meson mass and $\Delta
M=0.5\div 1.0$ GeV) split to the $cc$-baryon. However, it is known
\cite{8a} that the phenomenological application of the quark-hadron
duality method for heavy quarkonium photo- and hadroproduction needs
to introduce additional process dependent $K$-factor which is equal to
$1/3\div 1/6$ for description of the experimental data. Therefore
the results of Refs. \cite{6,7} may be considered as a rough
estimation of the double heavy baryon production rates, which need in
more detail analysis.

\section{The model}
In the present paper we consider a new mechanism of the double heavy
baryon production via heavy quark fragmentation based on the
hypothesis of the point-like diquarks, which can be produced directly
in the hard interactions of quarks and gluons at the short distance
\cite{8}. As it was shown in Ref. \cite{9} the data on $e^+e^-$
annihilation into hadrons don't contradict to the existence of the
very small diquark $(Qq)$ containing one heavy and one light quark.
In such a way, the heavy quark ($Q=b,c$) can fragment directly into
double heavy baryon spin-1/2 or -3/2 catching the scalar or the vector
heavy diquark ($D=(cq),\quad q=u,d,s$), correspondingly. In the
process of the production quark-diquark system $Q(Qq)$ in the colour
singlet state the relevant QCD scale satisfies to the next condition:
$\mu\ge m_Q+2m_D\gg\Lambda_{QCD}$, where $m_D$ is the heavy diquark
mass. This fact leads to the factorization of the double heavy baryon
production amplitude. The transition of the $Q(Qq)$ into final baryon
can be described using the nonrelativistic approximation of the
potential model, because the system $Q(Qq)$ contains two heavy
particles.

So, for the calculation of the baryon ($B=Q(Qq)$)
wave function at the origin we have used equation \cite{10}, which
effectively takes into account the relativistic effects of the
kinematic nature \cite{11}:
\begin{equation}
\left (-{\nabla^2\over{2\mu_R}}+U_{QD}(r)\right )\Psi(\vec
r)=E\Psi(\vec r),
\end{equation}
where
$$E={P^2\over{2\mu_R}}, \quad
\mu_R={M^2-(m_Q^2-m_D^2)^2\over{4M^3}},$$
$$P^2={[M^2-(m_Q+m_D)^2][M^2-(m_Q-m_D)^2]\over{4M^2}}.$$
The potential of the quark-diquark interaction is taken in the
conventional form \cite{3}:
\begin{equation}
U_{QD}=-{b\over r}+ar,
\end{equation}
where $a=0.183$ GeV$^2$, $b=0.52$. The results of our numerical
calculations for the radial part of the wave function at the origin
$|R(0)|^2=4\pi|\Psi(0)|^2$, are presented in Tables 1 and 2.

The fragmentation function $D_{Q\to B}(z,\mu)$ at the scale
$\mu=\mu_o= m_Q+2m_D$ for the production of the baryon $B$, containing
the heavy quark $Q$ and the heavy diquark $(Qq)$, is given by the
next expression:
\begin{equation}
D_{Q\to B}(z,\mu_o)={1\over{16\pi^2}}\int_{s_{min}}^{\infty}ds
\lim\limits_{q_o\to\infty}{\overline{|{\cal M}|^2}\over
{\overline{|{\cal M}_o|^2}}},
\end{equation}
where ${\cal M}$ is the matrix element for the production of a baryon
$B$ and antidiquark $\bar D$ with the total four-momentum $q$ and
the invariant mass $s=q^2$,
 ${\cal M}_o$ is the matrix element for the production of a quark $Q$
with the same three-momentum $\vec q$. The lower limit in the integral
(3) is
\begin{equation}
s_{min}={{M^2+\vec p^2_T}\over z}-{{m_D^2+\vec p^2_T}\over{1-z}}.
\end{equation}
Here $M=m_Q+m_D$ is the baryon mass, $\vec p_T$ is the baryon
transverse momentum in the reference frame where
$$q=(q_o,0,0,q_3),\quad p=(p_o,\vec p_T,p_3)\mbox{ and
}z={p_o+p_3\over{q_o+q_3}}.$$
In the limit of $q_o\to\infty$:
$$s_{min}={M^2\over z}-{m_D^2\over{1-z}}.$$

In the axial gauge associated with four-vector
$n=(1,0,0-1)$:
$$d_{\mu\nu}(k)=-g_{\mu\nu}+{k_{\mu}n_{\nu}+k_{\nu}n_{\mu}\over{(kn)}},$$
the fragmentation contribution comes only from the Feynman diagram for
$Q\to QD\bar D$ shown in Fig.1.

The gluon couplings to scalar and vector diquarks are presented by the
next expressions:
\begin{equation}
S_{\mu}^b=-ig_sT^b(q'-p_D)_{\mu}F_S(k^2),
\end{equation}
$$V_{\mu}^b =ig_sT^b \biggl \{ \varepsilon^*_D\varepsilon^*_{\bar D}
(q'-p_D)_{\mu}F_1(k^2)-
 [(q'\varepsilon^*_{\bar D})\varepsilon^*_{D\mu}-
(p_D\varepsilon^*_D)\varepsilon^*_{\bar D\mu}] F_2(k^2)-$$
\begin{equation}
(\varepsilon^*_D q')(\varepsilon^*_{\bar D} p_D)
(q'-p_D)_{\mu}F_3(k^2)\biggl \},
\end{equation}
where $T^b$ are Gell-Mann matrices,  $\varepsilon^*_D,
\varepsilon^*_{\bar D}$ are the diquark polarization vectors.
$F_s,F_1, F_2$ and $F_3$ are form factors depending on the momentum
transfer squared $k^2=(q'+p_D)^2$.

The light diquark form factors may be parametrized as follows
\cite{13,14}:
\begin{equation}
F_s(Q^2)={Q_s^2\over{Q_s^2+Q^2}}
\qquad
F_1(Q^2)=\Biggl({Q_v^2\over{Q_v^2+Q^2}}\Biggr)^2,
\end{equation}
$$F_2(Q^2)=(1+\kappa)F_1(Q^2),\qquad F_3(Q^2)=0$$
where $\kappa\approx 1.39$ being the anomalous magnetic moment
of the vector diquark, $Q_v^2\approx Q_s^2/2$ and
$Q_s^2\approx 3$ GeV$^2$ \cite{14}.

At present time the form factors of the heavy diquarks are unknown and
one has to assume a certain dependence on $Q^2$. The general consideration
(see, for example, Ref. \cite{9}), based on asymptotic QCD picture and
the phenomenology, predicts
$F_s\sim 1/Q^2,\quad F_1\sim F_s^2,\quad Q_s^2\gg 3$ GeV$^2$,
corresponding to a very small heavy diquark. We try to take these
points into account by using the simplest possible expressions for the
heavy diquark form factors:
$$F_s(Q^2)={Q_s^2\over Q^2},$$
\begin{equation}
F_1(Q^2)=\left ({Q_v^2\over Q^2}\right )^2,
\end{equation}
$$F_2(Q^2)=(1+\kappa)F_1(Q^2),\quad F_3(Q^2)=0,$$
where $Q_s^2=k_{min}^2=4m_D^2,\quad Q_v^2=Q_s^2/2$.
It has $k_{min}^2\approx 20$ GeV$^2$ for $(cq)$ diquarks and our
parametrization is the same as in Ref. \cite{9}.
Note, that such choice of the diquark form factors gives opportunity
to obtain the compact analytical expressions for the fragmentation
functions.


In the case of the heavy quark fragmentation into spin-1/2 baryon it
has fusion of the heavy quark $Q$ and scalar diquark $D$. After the
some obvious simplifications we have obtained:

\begin{equation}
{\cal M}_{1/2}=\Psi(0)\sqrt{M\over{2m_Qm_D}}
g_s^2{4\delta^{ij}\over{3\sqrt{3}}}{F_s(k^2)\over{(s-m_Q^2)^2}}
\end{equation}
$$
2\bar U(p_Q)\Biggl [-M(\hat
q+m_Q)+(s-m_Q^2){(np)\over{(nk)}}\Biggr ]\Gamma,
$$
where $g_s=\sqrt{4\pi\alpha_s}$,
$p_Q=\bar r p$ and $p_D=rp$ are the momenta of the quark and the
diquark in the baryon, correspondingly, $r=m_D/M,\quad \bar r=1-r$,
$p$ is the baryon momentum, $4\delta^{ij}/3\sqrt{3}$ is the colour
factor of the amplitude shown in Fig.1, the Dirac spinor $\Gamma$ is
the matrix element for the production of a $Q$ quark of momentum
$q=p+q'$, $q'$ is the antidiquark momentum, the spinor $\bar U(p)$
describes spin-1/2 baryon in the final state.

The amplitude for heavy quark fragmentation into spin-3/2 baryon,
corresponding to fusion of the heavy quark  and the vector diquark,
can be written as a sum of two parts, which are proportional to $F_1$
and $F_2$ form factors:
\begin{equation}
{\cal M}_{3/2}={\cal M}_{3/2}^1+{\cal M}_{3/2}^2,
\end{equation}

\begin{equation}
{\cal M}_{3/2}^1=\Psi(0)\sqrt{M\over{2m_Qm_D}}
g_s^2{4\delta^{ij}\over{3\sqrt{3}}}{F_1(k^2)\over{(s-m_Q^2)^2}}
\end{equation}
$$
2\bar\Psi_{\mu}(p_Q)\varepsilon^*_{\bar D\mu}
\Biggl [-M(\hat q+m_Q)+(s-m_Q^2){(np)\over{(nk)}}\Biggr ]\Gamma,
$$
\begin{equation}
{\cal M}_{3/2}^2=\Psi(0)\sqrt{M\over{2m_Qm_D}}
g_s^2{4\delta^{ij}\over{3\sqrt{3}}}{F_2(k^2)\over{(s-m_Q^2)^2}}
\end{equation}
$$
{1\over r}\bar\Psi_{\sigma}(p_Q)\varepsilon^*_{\bar D\lambda}
\Biggl
[k_{\sigma}\gamma_{\lambda}+{1\over{(kn)}}\Bigl(k_{\lambda}n_{\sigma}
\hat k-k_{\sigma}n_{\lambda}\hat k\Bigr)\Biggr ]
(\hat q+m_Q)\Gamma.
$$
Here, the spin-vector $\bar\Psi_{\mu}(p)$ describes the spin-3/2
baryon in the final state and satisfies to the following conditions
\cite{15}:
$$(\hat p-M)\Psi_{\mu}(p)=0,$$
$$\bar\Psi_{\mu}(p)\Psi_{\mu}(p)=2M,\quad \gamma_{\mu}\Psi_{\mu}=
p_{\mu}\Psi_{\mu}=0.$$
In the case of the unpolarized spin-3/2 baryon the summing on
the helicity states is carried out by means of the
following formula:
\begin{equation}
\sum_{\lambda}\Psi_{\mu}^{\lambda}(p)\bar\Psi_{\nu}^{\lambda}(p)=
(\hat p+M)
\biggl(g_{\mu\nu}-{1\over3}\gamma_{\mu}\gamma_{\nu}-{2p_{\mu}p_{\nu}
\over{3M^2}}+{{p_{\mu}\gamma_{\nu}-p_{\nu}\gamma_{\mu}}\over{3M}}\biggr).
\end{equation}

\section{The results}
Substituting the amplitudes (9) and (10) to the basic formula (3) we have
obtained the fragmentation functions
for a heavy quark split into the spin-1/2 and spin-3/2 baryons.
Omitting the details of the calculation  we write here final result for the
heavy quark fragmentation function for spin-1/2 baryon:
\begin{equation} D_{1/2}(z,\mu_o)=
{8\alpha_s^2(2m_D)\over
{405r^3}}{|\Psi(0)|^2\over{M^3}}{Q_s^4\over{M^4}} F_{1/2}(z,r),
\end{equation}
where
$$F_{1/2}(z,r)={z^4(1-z)^3\over
{(1-z+rz)^{10}}}\Biggl [15-6z(7+3r)+z^2(44+6r+15r^2)$$
$$-22z^3(1-r)+5z^4(1-r)^2\Biggr ].$$
The fragmentation probability for the production of the spin-1/2
baryon is
\begin{equation}
\int_0^1D_{1/2}(z,\mu_o)dz= {8\alpha_s^2(2m_D)\over
{405r^3}}{|\Psi(0)|^2\over{M^3}}{Q_s^4\over{M^4}} I_{1/2}(r),
\end{equation}
where
$$I_{1/2}(r)={1\over{42r^4(1-r)^{10}}}
[1-12r+75r^2-420r^3-1827r^4-126r^5+$$
$$2037r^6+300r^7-30r^8+2r^9
-1260r^4(1+r)^2\ln (r)].$$

The fragmentation function for $Q$ split into spin-3/2 baryon is
\begin{equation}
D_{3/2}(z,\mu_o)={\alpha_s(2m_D)\over{1215r^7}}{|\Psi(0)|^2\over{M^3}}
{Q_v^8\over{M^8}}F_{3/2}(z,r),
\end{equation}
where
$$F_{3/2}(z,r)={4\over 7}\Phi_{11}+4(1+\kappa)\Phi_{12}
+(1+\kappa)^2\Phi_{22},$$

$$\Phi_{11}(z,r)={z^4(1-z)^3\over{(1-z+rz)^{14}}}
  [7z^8(5r^6-18r^5+51r^4-76r^3+51r^2-18r+5)$$
      $$+2z^7(105r^5-455r^4+1122r^3-1038r^2+413r-147)
      +z^6(105r^6-42r^5+1596r^4-$$
$$2648r^3+ 5581r^2-2198r+1134)
     +2z^5(-21r^5-945r^4-100r^3-4352r^2+1505r$$
$$-1295)+z^4(847r^4+2028r^3+8031r^2-2170r+3780)+2z^3(-446r^3-$$
$$2018r^2+343r-1785)+7z^2(121r^2+2r+302)+42z(-r-17)+105],$$

$$\Phi_{22}(z,r)={z^4(1-z)^3\over{(1-z+rz)^{10}}}
[45-18z(7+3r)+z^2(164+84r+45r^2)-$$
$$2z^3(65+43r+15r^2)+z^4(47+56r+65r^2)],$$
$$\Phi_{12}(z,r)={z^4(1-z)^3\over{(1-z+rz)^{12}}}
[z^6(1-9r+17r^2-19r^3+10r^4)-$$
$$z^5(7-28r+35r^2-22r^3)+3z^4(11+r+22r^2+16r^3+10r^4)-$$
$$z^3(82+87r+95r^2+51r^3)+z^2(103+98r+47r^2)$$ $$-3z(21+11r)+15].$$

The fragmentation probability for the production of the spin-3/2
baryon can be calculated in the same way as for spin-1/2 baryon. We
don't present this result here because of it's unwieldy.

The fragmentation function  $D_{Q\to B}(z,\mu)$
satisfies to the Gribov-Lipatov-Altarelli-Parisi (GLAP) evolution
equations
\begin{equation}
\mu{\partial\over{\partial \mu}}D_{Q\to B}(z,\mu)= \int_z^1{dy\over
y}P_{Q\to Q}\left ({z\over y},\mu\right )D_{Q\to B}(y,\mu),
\end{equation}
where
$$P_{Q\to Q}(x,\mu)={4\alpha_s(\mu)\over{3\pi}}
\Biggl( {{1+x^2}\over{1-x}}\Biggr)_+,\quad f(x)_+=f(x)-\delta
(1-x)\int_0^1 f(x')dx'.$$
The boundary condition on the evolution equation is the initial
fragmentation function $D_{Q\to B}(z,\mu_o)$ at the scale
$\mu_o=m_Q+2m_D$. Note, that at leading order in $\alpha_s$ one has:
$$\int_o^1P_{Q\to Q}(z,\mu)dz=0,$$
and the evolution equation implies that the fragmentation probability\\
$\int_0^1D_{Q\to B}(z,\mu^2)$ does not evolve with the scale $\mu$.
Therefore the fragmentation probability is the universal characteristic
of the production rates. The evolution only changes the $z$-distribution
to smaller values of $z$.

The results of calculation for the fragmentation probabilities and
 the average values of the momentum fraction for $Q$ quark to split
into double heavy spin-1/2 and spin-3/2 baryons are shown in Tables 1
and 2. The values of $<z>$ are presented at $\mu=\mu_o$ and
$\mu=M_Z/2$. We used the following set of the mass parameters:
$m_c=1.7$ GeV, $m_b=5.1$ GeV, $m_{cu}^S=m_{cd}^S=1.9$ GeV,
$m_{cs}^S=2.0$ GeV. The masses of the vector diquarks are
greater than corresponding scalar diquarks by 0.1 GeV.
The fragmentation functions
$D_{c\to \Xi_{cc}}(z,\mu)$
and $D_{b\to\Lambda_{bc}}(z,\mu)$, divided by the fragmentation
probabilities, are shown in Fig. 2 at
 $\mu=\mu_o$ and $\mu=M_Z/2$.
In the range of accuracy of our model the shape of $z-$spectra for
spin-1/2 and spin-3/2 baryons are the same one to other as well as
$z-$spectra for baryons with strange quark
 ($\Omega_{cc},\Omega_{bc}$)  and without it ($\Xi_{cc},\Lambda_{bc}$).

Let  compare our results with the estimation \cite{7}, where it
was founded that fragmentation probabilities for heavy quark to split
into double heavy baryons are $(2\div 3)\cdot 10^{-5}$ independently on
the baryon spin and the flavour content. In our approach the fragmentation
probability depends on baryon spin and flavour.

Moreover, there is the relative growth of the production rates of the
spin-3/2 baryons for $b$-quark fragmentation in comparison with the
fragmentation of $c$-quark. We have obtained that
$\Xi_{cc}^*/\Xi_{cc}\sim 0.7$ and $\Lambda_{bc}^*/\Lambda_{bc}\sim
1.8$. Our results for the fragmentation probabilities for the spin-3/2
baryons are approximately equal to the results of \cite{7}, however
the fragmentation probabilities for spin-1/2 baryons are smaller about
factor 2. Taking into consideration the uncertainties of our
calculation, connected with the diquark form factors (8), we can
predict the fragmentation probabilities for the $cc-$ and $bc-$
baryons are about of $10^{-5}$. There are more accurate predictions
for baryon $z$-spectra in the our approach, which are practically
independent on the diquark form factors and other parameters. We have
obtained that the average fraction of the baryon momentum $<z>\approx
0.54$ for the $cc-$baryons and $<z>\approx 0.66$ for the $bc-$baryons
at $\mu=M_Z/2$, and $<z>\approx 0.75$ and  $<z>\approx 0.84$,
correspondingly, at $\mu=\mu_o$. This results are independent from the
baryon spin. By contrast, in the model, based on $Q\to (QQ)$
fragmentation mechanism corresponding values at $\mu=\mu_o$ are equal:
$<z>=0.57\div 0.62$ for the $cc-$baryons and $<z>=0.68\div 0.73$ for
the $bc-$baryons.

In conclusion, we sum the results of our paper. In the framework of NR
QCD and a quark-diquark model of baryons we have obtained in the
leading order in $\alpha_s$ the fragmentation functions and
the probabilities for heavy quark to split into double heavy spin-1/2 and
spin-3/2 baryons. Using the QCD evolution equations we have recalculated
the fragmentation functions from the initial scale $\mu_o$ to
$\mu=M_Z/2$. Our results can be used for prediction of the double heavy
baryon production rates as well as for the description of the energy
spectra at $e^+e^-$-collider LEP. Contrary to Refs. \cite{6,7} we have
predicted also	few precise effects: the nontrivial spin and flavour
dependence of the baryon production rates, the value of the average
baryon momentum $<z>$ and it's dependence from the scale $\mu$ as well
as from the baryon flavours.

The authors express their gratitude to R.N.\,Faustov, V.O.\,Galkin,
V.V.\,Kiselev, and A.K.\,Likhoded for the discussions of the quarkonium
physics as well as C.\,Carimalo and D.B.\,Lichtenberg for the
information on diquark models.


\vspace{10mm}


\begin{center} Table 1. The spin-1/2 baryons.
\vspace{4mm}

\begin{tabular}{||c|c|c|c|c|c||}\hline
Baryon& M, & $|R(0)|^2$, &
$P_{Q\to B}$& $<z>_o$& $<z>_{M_Z/2}$ \\
 &[GeV]& [GeV$^3$]& & &\\ \hline\hline
$\Xi_{cc}$& 3.6& 1.2& $5.8\cdot
10^{-6}$& 0.75& 0.55\\ \hline
$\Omega_{cc}$& 3.7& 1.3& $5.3\cdot
10^{-6}$& 0.74& 0.54\\ \hline
$\Lambda_{bc}$& 7.0& 3.8& $8.9\cdot
10^{-6}$& 0.84& 0.67\\ \hline
$\Omega_{bc}$& 7.1& 4.0& $7.4\cdot
10^{-6}$& 0.83& 0.66\\ \hline
\end{tabular}
\vspace{5mm}

Table 2. The spin-3/2 baryons.
\vspace{4mm}

\begin{tabular}{||c|c|c|c|c|c||}\hline
Baryon& M, & $|R(0)|^2$, &
$P_{Q\to B}$& $<z>_o$& $<z>_{M_Z/2}$ \\
 &[GeV]& [GeV$^3$]& & &\\ \hline\hline
$\Xi_{cc}^*$& 3.7& 1.25&
$1.2\cdot 10^{-5}$& 0.75& 0.54\\ \hline
$\Omega_{cc}^*$& 3.8& 1.35&
$1.1\cdot 10^{-5}$& 0.74& 0.53\\ \hline
$\Lambda_{bc}^*$& 7.1& 4.0&
$3.5\cdot 10^{-5}$& 0.84& 0.67\\ \hline
$\Omega_{bc}^*$& 7.2& 4.1&
$3.3\cdot 10^{-5}$& 0.83& 0.66\\ \hline
\end{tabular}
\end{center}

\vspace{10mm}

\begin{center}
Figure captions.
\end{center}

1. Diagram used for description of the heavy quark to split into
double heavy baryon.

2. The fragmentation functions normalized to the unity at $\mu=\mu_o$
(curves 1 and 3) and $\mu=M_Z/2$ (curves 2 and 4). The curves 1 and 2 -
$D_{\Xi_{cc}}(z,\mu)$, the curves 3 and 4 - $D_{\Lambda_{bc}}(z,\mu)$.

\pagebreak

\def\emline#1#2#3#4#5#6{%
       \put(#1,#2){\special{em:moveto}}%
       \put(#4,#5){\special{em:lineto}}}

{ ~ }

\vspace{3cm}

\begin{center}
\unitlength=1mm
\special{em:linewidth 1pt}
\linethickness{1pt}
\begin{picture}(75.00,47.00)
\put(20.00,20.00){\circle{10.00}}
\emline{25.00}{20.00}{1}{40.00}{20.00}{2}
\emline{40.00}{20.00}{3}{50.00}{33.00}{4}
\emline{50.00}{33.00}{5}{50.00}{33.00}{6}
\emline{59.00}{20.00}{7}{68.00}{33.00}{8}
\emline{59.00}{20.00}{9}{75.00}{20.00}{10}
\put(59.00,32.50){\oval(18.00,3.00)[]}
\emline{65.00}{31.00}{11}{57.00}{20.00}{12}
\emline{57.00}{18.00}{13}{75.00}{18.00}{14}
\put(58.00,19.00){\circle*{4.00}}
\emline{40.00}{20.00}{15}{43.00}{20.00}{16}
\emline{45.00}{20.00}{17}{48.00}{20.00}{18}
\emline{50.00}{20.00}{19}{53.00}{20.00}{20}
\emline{55.00}{20.00}{21}{57.00}{20.00}{22}
\put(20.00,20.00){\makebox(0,0)[cc]{$\Gamma$}}
\emline{57.00}{34.00}{23}{66.00}{45.00}{24}
\emline{60.00}{34.00}{25}{68.00}{44.00}{26}
\emline{63.00}{44.00}{27}{69.00}{46.00}{28}
\emline{69.00}{46.00}{29}{67.00}{40.00}{30}
\put(72.00,47.00){\makebox(0,0)[cc]{$p$}}
\put(72.00,12.00){\makebox(0,0)[cc]{$q'$}}
\put(34.00,12.00){\makebox(0,0)[cc]{$q$}}
\put(39.00,28.00){\makebox(0,0)[cc]{$p_Q$}}
\put(70.00,28.00){\makebox(0,0)[cc]{$p_D$}}
\end{picture}
\end{center}

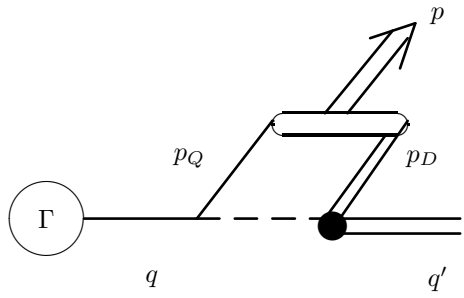
\begin{figure}[h]
\caption{ }
\end{figure}
\newpage
%
%

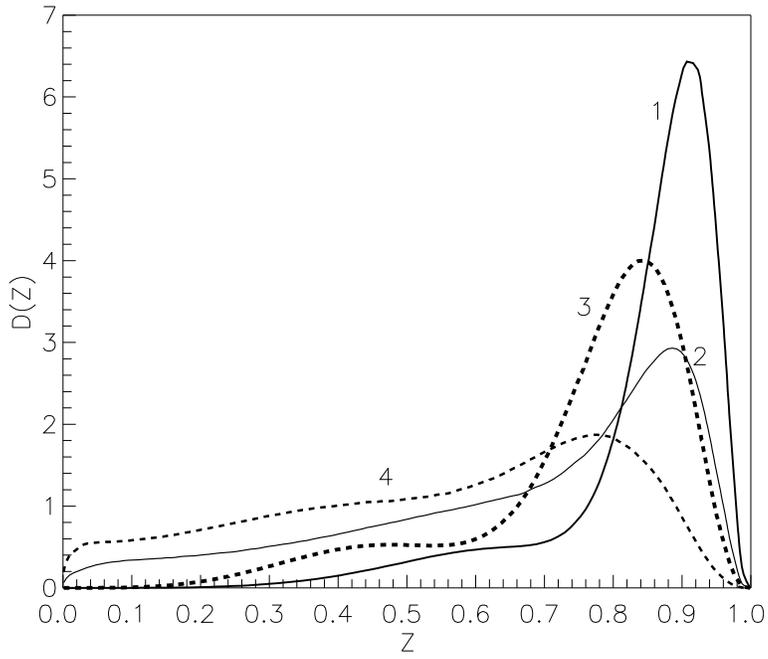
\end{document}